\documentclass[aps,prb,superscriptaddress,twocolumn,floatfix]{revtex4-1}

\usepackage{graphicx,color}
\usepackage{dcolumn}
\usepackage{bm}
\usepackage{stmaryrd}
\usepackage{latexsym}
\usepackage{amssymb}
\usepackage{amsfonts}
\usepackage{amsmath}
\usepackage{subfigure}
\usepackage{hyperref}
\usepackage{verbatim}
\usepackage{multirow}

\begin{document}

\preprint{APS/123-QED}

\title{Renormalization of the Mott gap by lattice entropy: The case of 1T-TaS$_2$}
\author{Li Cheng}
\affiliation{School of Mathematical and Physical Sciences, Dalian University of Technology, Panjin 124221, China}
\affiliation{Institute for Advanced Study, Tsinghua University, Beijing 100084, China}
\author{Shunhong Zhang}
\affiliation{International Center for Quantum Design of Functional Materials (ICQD), Hefei National Laboratory for Physical Sciences at the Microscale, and CAS Center For Excellence in Quantum Information and Quantum Physics, University of Science and Technology of China, Hefei, Anhui 230026, China}
\author{Shuang Qiao}
\affiliation{Beijing Computational Science Research Center, Beijing 100193, China}
\author{Xiaofeng Wang}
\affiliation{School of Mathematical and Physical Sciences, Dalian University of Technology, Panjin 124221, China}
\author{Lizhao Liu}
\email{lizhao$_$liu@dlut.edu.cn}
\affiliation{School of Mathematical and Physical Sciences, Dalian University of Technology, Panjin 124221, China}
\affiliation{Key Laboratory of Materials Modification by Laser, Ion and Electron Beams (Dalian University of Technology), Ministry of Education, Dalian 116024, China}
\author{Zheng Liu}
\email{zheng-liu@tsinghua.edu.cn}
\affiliation{Institute for Advanced Study, Tsinghua University, Beijing 100084, China}
\affiliation{State Key Laboratory of Low-Dimensional Quantum Physics, Department of Physics, Tsinghua University, Beijing 100084, China}
\date{\today}

\begin{abstract}
In many transition-metal oxides and dichalcogenides, the electronic and lattice degrees of freedom are strongly coupled, giving rise to remarkable phenomena, such as metal-insulator transition (MIT) and charge-density wave (CDW) order. We study this interplay by tracing the instant electronic structure under \textit{ab initio} molecular dynamics. Applying this method to a 1T-TaS$_2$ layer, we show that the CDW-triggered Mott gap undergoes a continuous reduction as the lattice temperature raises, despite a nearly constant CDW amplitude. Before the CDW order undergoes a sharp first-order transition around the room temperature, the dynamical CDW fluctuation already shrinks the Mott gap size by half. The gap size reduction is one order of magnitude larger than the lattice temperature variation. Our calculation not only  provides an important clue to understand the thermodynamics behavior in 1T-TaS$_2$, but also demonstrates a general approach to quantify the lattice entropy effect in MIT.

\end{abstract}
\maketitle

\section{Introduction}
1T-TaS$_2$ has perhaps the richest electronic phase diagram of all transition-metal dichalcogenides because of the intertwined lattice, charge, orbital and spin degrees of freedom \cite{pressure2008}. While the low-temperature commensurate charge density wave (CCDW) order and the accompanied metal-to-insulator transition (MIT) have been investigated for a long time by diffraction \cite{APS_1975}, transport \cite{PMB_1979}, scanning tunneling microscopy (STM) \cite{Sci_1989,PRL_1991, PRL1994} and angle-resolved photoemission \cite{PRB1981,smith1985band,manzke1988electronic}, the absence of magnetic susceptibility of the insulating phase remains puzzling \cite{EPL1989}. The possibility of a quantum spin liquid state due to the lattice frustration was proposed recently \cite{lee2017}, which aroused revived theoretical interest and stimulated a series of recent experiments \cite{2017high,npj2017,gapless2017,PRB2017metallic,PRB2018real}.

\begin{figure*}
\centering
\includegraphics[width=18 cm]{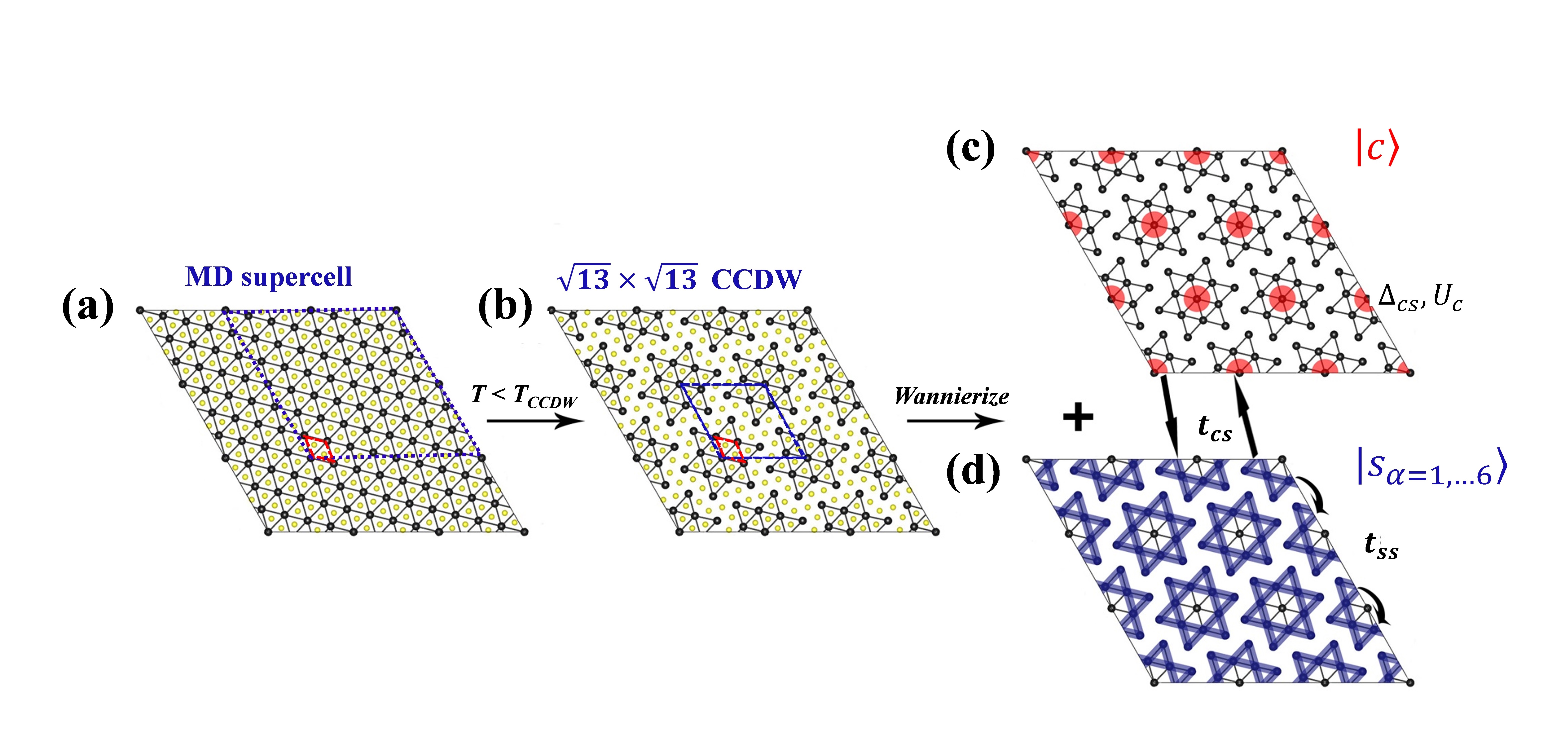}
\caption{Schematics of (a) the high-temperature lattice of  a 1T-TaS$_2$ layer; (b) the low-temperature CCDW structure; (c)(d) differentiation of two types of low-energy Wannier orbitals by CCDW, as characterized in Ref. \cite{qiao2017}. The  physical meaning of the parameters associated with the Wannier orbitals is discussed in Sec. \ref{wannier}.  
}
\label{orbital}
\end{figure*}

The general consensus \cite{Rossnagel_2011} is that below $ 200\ \pm $ 20 K ($T_{CCDW}$), the $\sqrt{13}\times\sqrt{13}$ CCDW order is fully established [Figs. \ref{orbital}(a)(b)]. The Ta atoms are grouped into 13-atoms clusters with a ``Star-of-David'' (SD) arrangement. It is widely perceived that such a 2D layer can be viewed as a cluster Mott insulator - each SD acts effectively as a correlated site with an odd number of electrons and the SDs form a triangular superlattice. Above $T_{CCDW}$ to around 350 K ($T_{NC}$), the so-called nearly CCDW phase emerges, consisting of a mixture of 
the SDs and discommensurate areas. Above $T_{NC}$, the SD clusters completely disappear, leaving a weak incommensurate CDW order. 

The Mott phase in 1T-TaS$_2$ features: (i) geometry frustration; (ii) a soft gap of the order of O(10$^2$) meV ~\cite{qiao2017,PRB_2015_cho,PRB_2018_lutsyk}; and (iii) the accompanied lattice distortion. Current investigations largely concentrate on the first two aspects. The first aspect serves as the basis to discuss the quantum spin liquid physics \cite{lee2017, 2017high,npj2017,gapless2017,PRB2017metallic,PRB2018real}, and the second renders various ways to control the MIT, e.g via pressurization\cite{pressure2008}, doping \cite{PRL1998,PRB_2016_shao,PRL2012,li2012fe,liu2013}, ionic liquid gating \cite{NatNano_Zhang}, voltage pulsing \cite{NatNano_Zhang, nano2016} and likely layer stacking \cite{Natphys_Geck}. 

This article aims to highlight the significance of the last aspect. The strong coupling between the electron and lattice degrees of freedom underlies many useful applications of transition metal oxides. For these systems, it has been shown that both lattice energy \cite{PRL18, PNAS19} and entropy \cite{nature14, nature19} have novel consequences. Similarly, it is reasonable to expect that lattice dynamics also plays an important role in 1T-TaS$_2$. By performing \textit{ab initio} molecular dynamics (MD) simulation, we first show that the first-order CCDW transition and the associated MIT can be reasonably reproduced as a function of the lattice temperature. Furthermore, we show that below the transition temperature, a continuous variation of the electronic band gap persists. 

\section{Computational method}
\subsection{Electronic ground state}
Born-Oppenheimer (BO) approximation is presumed throughout this article. For a given lattice structure, the electronic ground state is calculated within the framework of density functional theory plus on-site U correction, by using the Vienna \textit{ab initio} Simulation Package (VASP) \cite{kresse1996,kresse1996efficient}. 

The simulation cell contains a single layer of 52 Ta atoms sandwiched by 104 S atoms (in total $N_{atom}$ = 156), which can accommodate up to 4 SDs. A 15 $\AA$ vacuum layer is included in the z-direction.   

We employ the projector augmented wave method \cite{kresse1999ultrasoft} and the exchange-correlation functional due to Perdew, Burke and Ernzerhof (PBE) \cite{perdew1996}. The +U correction is employed to capture the Coulomb interaction of Ta 5d orbitals on the Hartree-Fock level, following the simplified (rotational invariant) approach introduced by Dudarev et al. \cite{U1998}. We employ an effective U = 2.27 eV as previously derived from the linear-response calculation \cite{PRB2014Darancet}. We use a plane-wave cutoff of 300 eV, and the Brillouin zone was sampled with the $\Gamma$ point only. The initial spin polarization of the 4 SDs in the simulation cell is set to be the same.

\subsection{Lattice dynamics}
Nuclei are subject to the Newton's equation of motion on the BO potential surface using a time step of 2 fs.  To simulate a canonical ensemble, the in-plane lattice constant is fixed to the experimental value \cite{Rossnagel_2011}, and the Nose thermostat is used to adjust the lattice temperature \cite{nose1984molecular,nose1984unified,nose1992}. For each T, the MD simulation lasts for 20 ps, and the last 4 ps is used to calculate thermodynamics properties, such as the equilibrium lattice structure and the CCDW order parameter.

Numerically, it is important to guarantee that the last 4 ps has already achieved thermal equilibrium. For a better convergence, we start from low T, which is closest to the DFT relaxed static structure. Then, the structure of the final MD step is used as the initial structure of the next temperature, which is elevated progressively. When all the SDs melt, we reversely reduce the temperature progressively, using the equilibrium structure at the higher temperature as input. Finally, the simulation forms a complete heating-cooling cycle. Our criterion for thermal equilibrium are that (i) the temperature fluctuation has already converged to $\sqrt{2/{(3N_{atom})}}=6.5\%$ as expected from a Boltzmann distribution, (ii) clear periodicity with constant amplitude can be observed from the atomic displacement, and (iii) the observables from the heating and cooling processes coincide when the temperature is away from the transition point. 

\subsection{Wannier function analysis}\label{wannier}
Maximally localized Wannier function (MLWF) analysis is used to determine the key electronic parameter coupled to the lattice degree of freedom, by using the WANNIER90 code \cite{W90}.  Based on the MLWF transformation, the single-electron band structure at the DFT-PBE level is mapped to a tight-binding model.

For the static lattice structure at 0 K, two types of MLWFs were characterized in our previous study, which sucessfully reproduced the orbital textures revealed by STM dI/dV maps ~\cite{qiao2017}. One type of MLWFs is associated with the central Ta atom of a SD ($|c\rangle$), which suffers from a strong onsite Coulomb repulsion, and the other consists of six hybridized orbitals along the edges of the SD ($s_{\alpha=1,...,6}$), which are much more delocalized. A schematic summary is shown in Figs. \ref{orbital}(c)(d). These seven orbitals in together accommodate 13 unpaired Ta d-electrons, with the topmost band half filled. The corresponding tight-binding model takes the form:
\begin{eqnarray}\label{eq:multi}
H_{hop}=\Delta_{cs}\sum_i c_i^{\dag}c_i&+&t_{cs}\sum_{i\alpha}(c_i^\dag s_{i\alpha}+h.c.) \nonumber \\
&+&\sum_{i\alpha,j\beta}t_{ss}^{i\alpha,j\beta}s_{i\alpha}^\dag s_{j\beta}, 
\end{eqnarray}
in which $c_i^\dag$ and $s_{i\alpha}^\dag$ are the creation operators of $|c\rangle$ and $|s_\alpha\rangle$ in the SD labeled by $i$. $\Delta_{cs}$ is the onsite energy difference between $|c\rangle$ and $|s_\alpha\rangle$, and $t_{cs}$ ($t_{ss}^{i\alpha,j\beta}$) are the hopping amplitude between the central and surrounding orbitals (two surrounding orbitals). By further including the interaction terms, in particular a strong onsite repulsion $U_c$ associated with $|c_i\rangle$, this Hamiltonian is expected to capture the low-energy electronic degrees of freedom.

Following the same recipe,  we extend the analysis to instant lattice structures during the MD simulation. Since the MD supercell contains four SDs in total, the total number of MLWFs is 28, including 4 $|c_i\rangle$ orbitals and 24 $|s_{i\alpha}\rangle$ orbitals. For all the extracted parameters, an average of the four SDs in the simulation cell is performed. 

\subsection{Validity and limitations}
The MD simulation is expected to nicely describe the lattice thermodynamics. Under equilibrium, the MD time average directly measures the ensemble average.  The most important lattice vibration that melts the SDs has the $\sqrt13\times\sqrt13$ wave vector, which is fully accommodated in our simulation cell. The phase transition naturally occurs when the kinetic energy of the atoms becomes large enough to escape the $\sqrt{13}\times\sqrt{13}$ potential well. 

The complexities of the stacking order of the layers in a 3D bulk and the interlayer coupling are beyond the scope of the current calculation. A finite-temperature phase transition in our 2D simulation does not violate the Mermin-Wagner theorem, because the imposed periodic boundary condition cuts off any thermal fluctuation with a wave vector larger than the cell size. However, we should  note that our simulation cell (see Fig. \ref{struct} ) is still not large enough to describe phase separation and long-wave incommensurate CDW. Experimentally, between $T_{CCDW}$ (the SDs start to melt) and $T_{NC}$ (the SDs completely melt), there is a wide range in which the CCDW domains and the discommensurate regions coexist \cite{Rossnagel_2011}. Our simulated transition temperature turns out to fall between the experimental $T_{CCDW}$ and $T_{NC}$.

It is understood that DFT+U is a mean-field symmetry-breaking approximation to the Mott insulating ground state. At 5 K, the DFT+U density of states achieved good agreement with the STM dI/dV spectrum \cite{qiao2017}. However, the main limitation is that electronic entropy is missing.  Rigorously, the simulated system should be viewed as a hypothetical one with lattice temperature only, while the electronic temperature is always zero. Combining a more advanced algorithm, e.g. dynamical mean-field theory \cite{RMP_DMFT},  with the lattice dynamics is currently beyond the computational capability. In general, an accurate description of the finite-temperature charge and spin fluctuations in a Mott insulator remains a theoretical challenge. Some discussions on the consequences of electronic temperature and a revisit of related experimental results are given in Sec. \ref{discussion}. 

\section{Results}

\begin{figure}
\centering
\includegraphics[width=8cm]{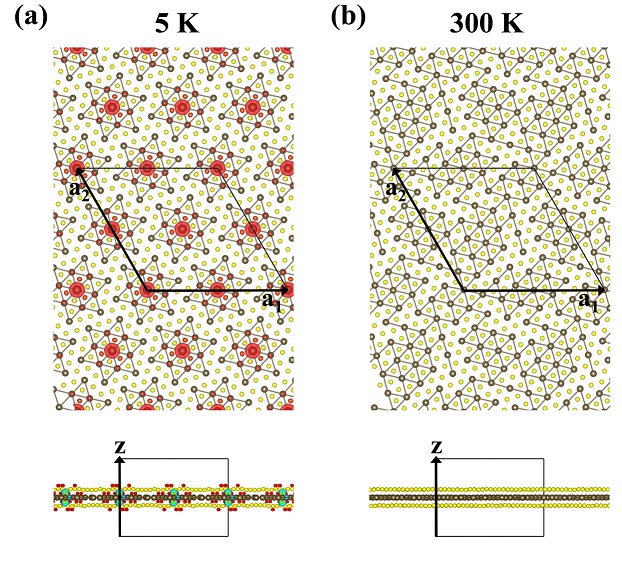}
\caption{Equilibrium lattice structures obtained from MD at (a) 5 K and (b) 300 K. The black box indicates the simulation cell and $a_1$, $a_2$, $z$ are the three cell vectors. The colored surface gives the spin density isovalue contour. A Ta-Ta bond is drawn when the Ta-Ta distance is below 3.4 $\AA$. }
\label{struct}
\end{figure}

The MD obtained equilibrium lattice structures at 5 K and 300 K are shown in Fig. \ref{struct} . The lattice temperature effect on the CCDW order can be clearly observed. The animation files of the equilibrium lattice dynamics at four typical temperatures are provided in the Supplemantary Materials.

\begin{figure}
\centering
\includegraphics[width=8.5cm]{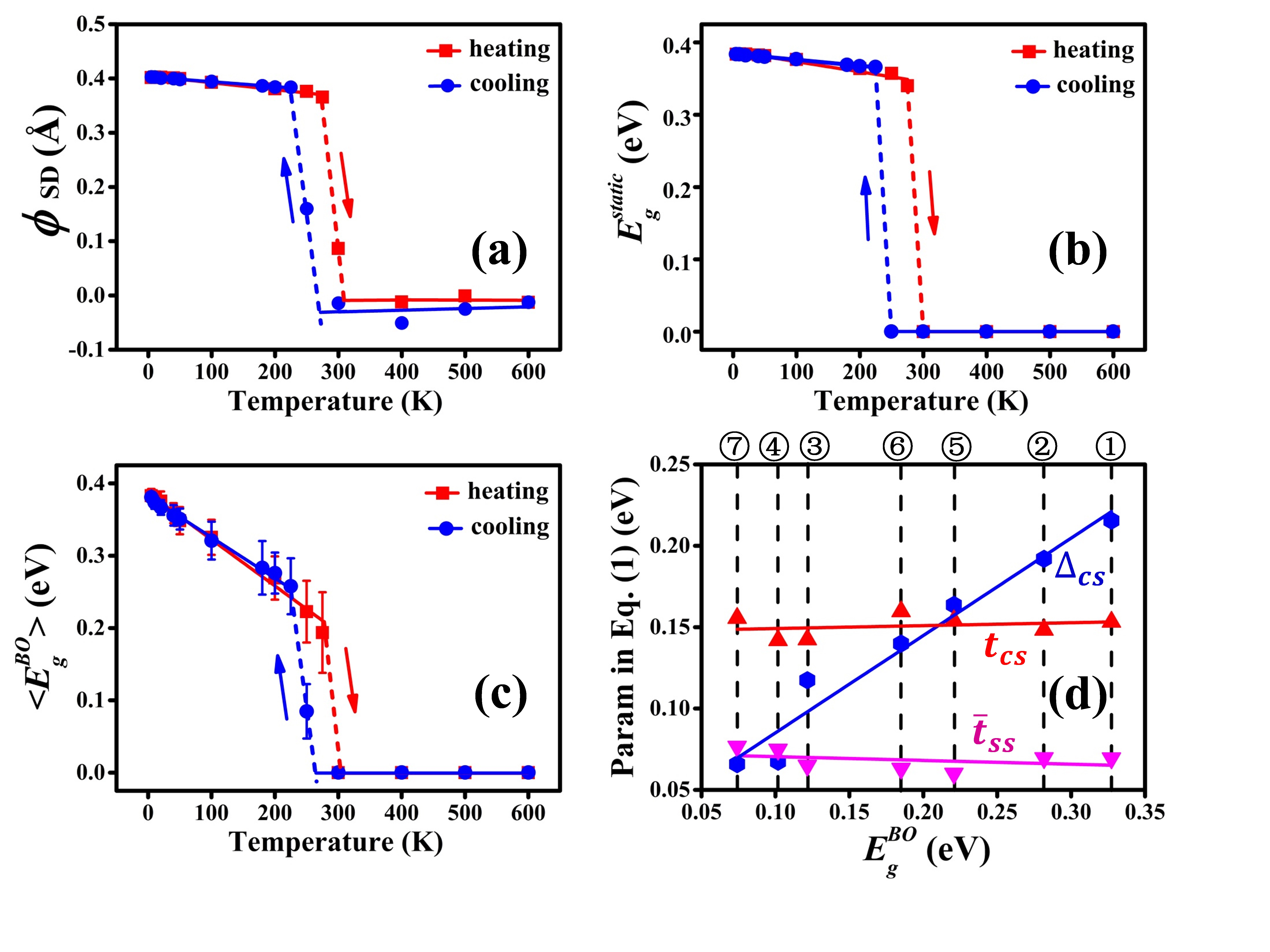}
\caption{(a-c) The temperature dependence of (a) the CCDW order parameter $\phi_{SD}$; (b) the static DFT+U(=2.27 eV) band gap $E_g^{static}$;  and (c) the time-averaged Born-Oppenheimer band gap $\langle E_g^{BO} \rangle$. The error bar in (c) is calculated from the standard deviation of the instant $E_g^{BO}$.  (d) Correlation between $E_g^{BO}$ and effective parameters in Eq. \ref{eq:multi} from MLWF analysis. The numbers on top denote the seven time slices marked in Figs. \ref{band}(e-g).}
\label{phase}
\end{figure}

We can define a CCDW order parameter $\phi_{SD}=\bar{d}_{inter}-\bar{d}_{intra}$, where $\bar{d}_{inter}$ ($\bar{d}_{intra}$) is the time-averaged Ta-Ta distance between SDs (within a SD). The definition of inter- (intra-)SD bonds is ambiguous in the high-T phase, so we always refer to the SD positions in the CCDW phase. Both $\bar{d}_{inter}$ and $\bar{d}_{intra}$ are determined from the equilibrium lattice structure as a function of T. When SDs  melt, $\phi_{SD}$ vanishes.  Figure \ref{phase}(a) plots $\phi_{SD}$ versus temperature. A sharp first-order transition can be observed. 
The transition temperature $T_C$ is 250 K $\sim$ 300 K.

\begin{figure*}
\centering
\includegraphics[width=18cm]{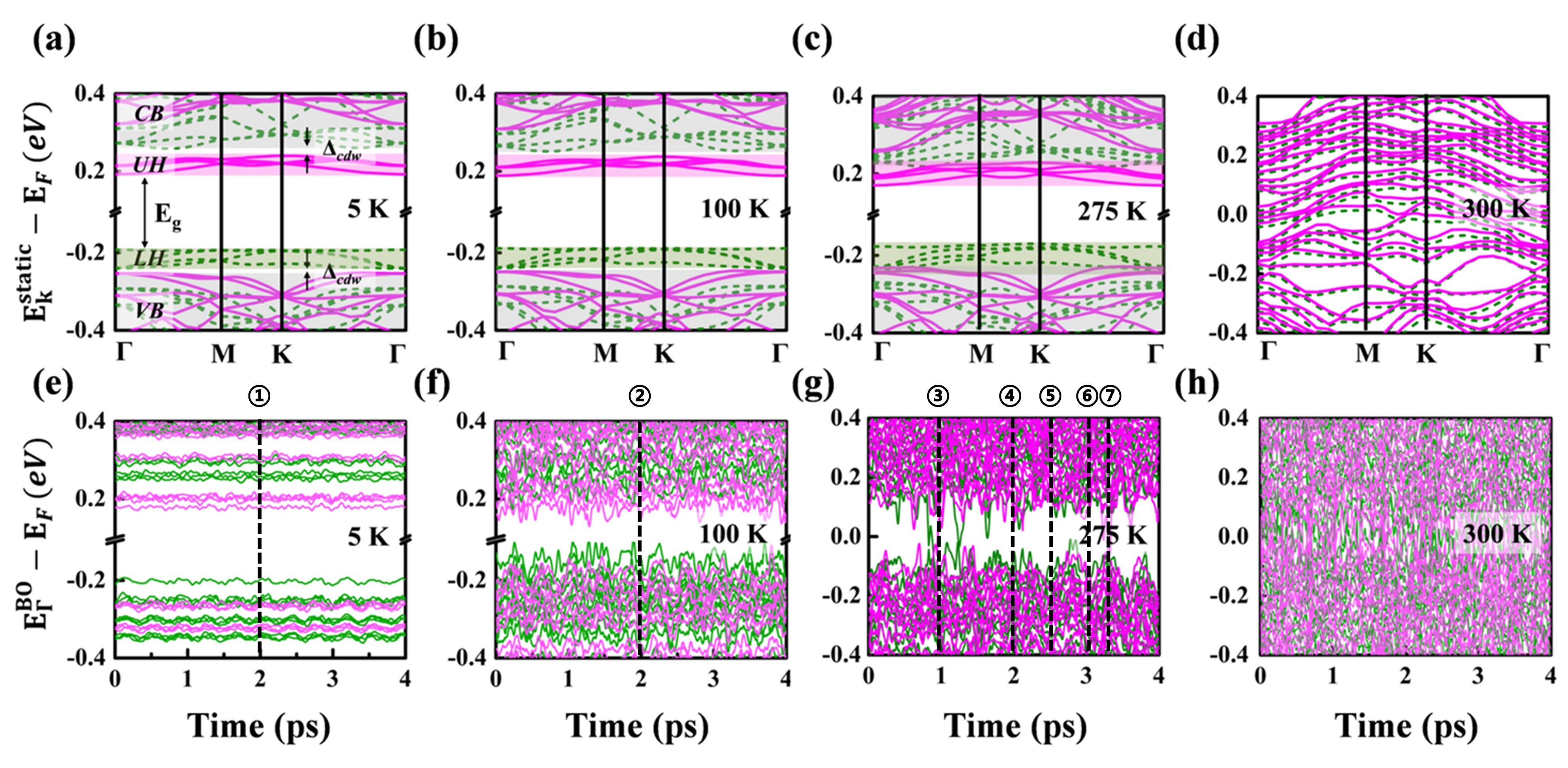}
\caption{ (a-d) Static DFT+U(=2.27 eV) band structures calculated from the time-averaged lattice structures at different temperatures. (e-h) Time evolution of the instant energy levels at the $\Gamma$ point within the last 4 ps of the MD simulation. The vertical dashed lines mark seven time slices used to extract the data in Fig. \ref{phase}(d). The pink and green bands indicate different spins.} 
\label{band}
\end{figure*}

Figure \ref{phase}(b) plots the static DFT+U band gap $E_g^{static}$ with respect to the equilibrium lattice structures at different lattice temperatures. A sharp metal-to-insulator transition takes place concurrently with the CCDW transition. Figs. \ref{band}(a-d) plot the static band structures at four typical lattice temperatures, which only show marginal dependence on the lattice temperature below T$_C$. 

Figures \ref{band}(e-h) plot the instant energy levels at the $\Gamma$ point as a function of MD duration. At 5 K, the gap size agrees well with the static band structure, despite slight temporal fluctuations. At higher temperatures, the temporal fluctuations significantly reduces the static gap size. At 275 K, instant level crossings can be observed, indicating that the system is close to the phase transition. The time-averaged band gap $\langle E_g^{BO} \rangle $ as a function of the lattice temperature is plotted in Fig. \ref{phase}(c). The gap size decreases by half, from around 0.4 eV at 5 K to around 0.2 eV at T$_C$ . It is remarkable that $\Delta\langle E_g^{BO} \rangle / k_B\Delta T$ is of the order of $-10$, indicating that a small variation of the lattice kinetic energy can result in a huge impact on the electronic structure. 

Seven instant structures are picked from Figs. \ref{band} (e-g), and the single-electron parameters in Eq. (\ref{eq:multi}) are extracted via Wannerisation. In Tab. \ref{table-1}, we list the obtained values of $\Delta_{cs}$, the nearest-neighbor  $t_{cs}$ and the six dominating $t_{ss}$'s as visualized in Fig. 7 of Ref. \cite{qiao2017}.  By plotting them against  $E_g^{BO}$ in Fig. \ref{phase}(d), a clear correlation between $\Delta_{cs}$ and $E_g^{BO}$ can be observed, while the hopping parameters do not play an active role. Note that the plotted $\bar{t}_{ss}$ is an average of the six $t_{ss}$'s in Tab. \ref{table-1}.

\begin{table}
\caption{Effective parameters in Eq. \ref{eq:multi} from MLWF analysis. The first column contains the previous data obtained from the fully-relaxed structure\cite{qiao2017}. The other seven columns correspond to the seven instant MD lattice structures marked in Figs. \ref{band}(e-g).}
\setlength\extrarowheight{5pt}
\begin{ruledtabular}
\begin{tabular}{ccccccccc}
Unit$ (eV) $ & [\cite{qiao2017}] & \textcircled{1} & \textcircled{2} & \textcircled{3} & \textcircled{4} & \textcircled{5} & \textcircled{6} & \textcircled{7}\\
\hline
\multirow {1}*{$\Delta_{cs}$} 
& \color{blue}0.212	& 0.215	& 0.192	& 0.117	& 0.067	& 0.164	& 0.140	& 0.066 \\
\multirow {1}*{$t_{cs}$}
& \color{blue}0.162	& 0.153	& 0.148	& 0.142	& 0.142	& 0.154	& 0.160	& 0.156 \\
\multirow {1}*{$t_{ss1}$}
& \color{blue}0.150	& 0.141	& 0.155	& 0.111	& 0.183	& 0.131	& 0.118	& 0.200 \\
\multirow {1}*{$t_{ss2}$}
& \color{blue}0.091	& 0.093	& 0.092	& 0.092	& 0.093	& 0.081	& 0.094	& 0.094 \\
\multirow {1}*{$t_{ss3}$}
& \color{blue}0.072	& 0.063	& 0.064	& 0.073	& 0.071	& 0.060	& 0.065	& 0.068 \\
\multirow {1}*{$t_{ss4}$}
& \color{blue}0.050	& 0.052	& 0.048	& 0.052	& 0.048	& 0.043	& 0.048	& 0.045 \\
\multirow {1}*{$t_{ss5}$}
& \color{blue}0.042	& 0.038	& 0.035	& 0.037	& 0.032	& 0.027	& 0.032	& 0.031 \\
\multirow {1}*{$t_{ss6}$}
& \color{blue}0.042	& 0.030	& 0.024	& 0.026	& 0.023	& 0.018	& 0.023	& 0.023 \\
\end{tabular}
\end{ruledtabular}
\label{table-1}
\end{table}

\section{Discussions}\label{discussion}

\begin{figure}
\centering
\includegraphics[width=9cm]{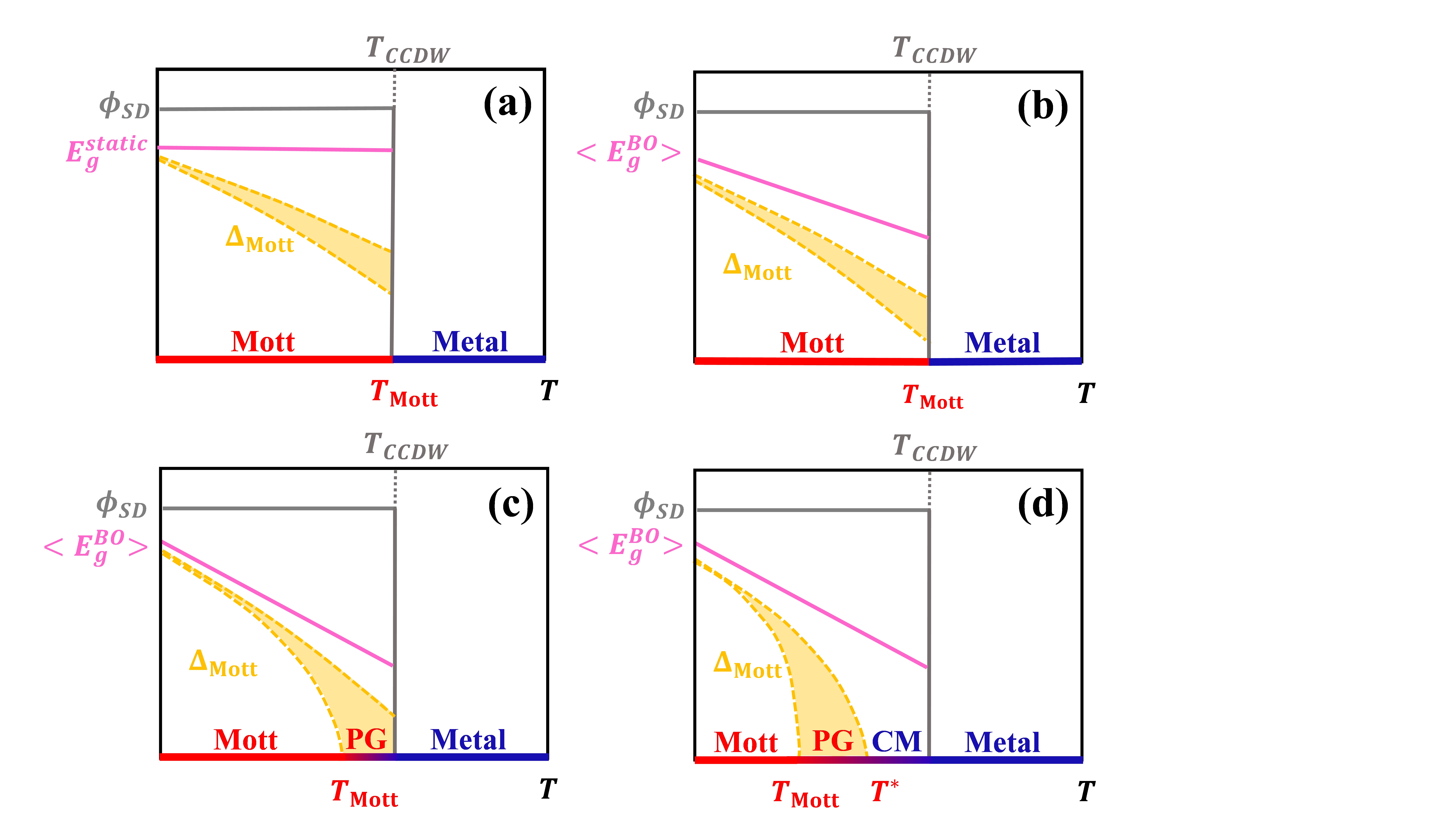}
\caption{Schematic summary of $\phi_{SD}(T)$, $E_g^{static}(T)$ and $E_g^{BO}(T)$. $\Delta_{Mott}$ is the speculated behavior of the realistic Mott gap when electronic entropy is also present. (a) reflects a scenario without considering the lattice entropy effect. (b)(c)(d) are possible outcomes when both the lattice and electronic entropy are present. In (c) and (d), ``PG'' denotes a pseudogap phase, in which the Mott gap $\Delta_{Mott}$ partially melts. ``CM'' denotes a correlated metal.}
\label{transition}
\end{figure}

The comparison between $E_g^{static}$ and $\langle E_g^{BO} \rangle$ clearly indicates that the driving force is not the static CCDW amplitude but the dynamical vibrations. This is in general related to the strong electron-phonon coupling (EPC), as expected from the CDW formation. 

Specifically, the Wannier function analysis reveals $\Delta_{cs}$ as the key parameter strongly coupled to phonons.  The central role of  $\Delta_{cs}$ during the MIT in 1T-TaS$_2$ coincides with the site-selective Mott transition scenario originally proposed for the rare-earth nickelates \cite{PRL12}. Namely, an on-site potential difference ($\Delta_{cs}$)  associated with a lattice distortion ($\phi_{SD}$) leads to a site-selective ($|c_i\rangle$) localization. The new insight from the MD simulation is that even when the mean value of $\phi_{SD}$ is fixed, its fluctuation can still result in giant renormalization of the electronic gap. 

It is difficult to quantify the electronic temperature effect in a Mott insulator of such a complicated supercell, not to mention in combination with lattice dynamics. Unlike in a conventional semiconductor, the electronic excitations in a Mott insulator go beyond a plain band scenario, and in particular the spin fluctuations play an important role \cite{RMP_Wen}. According to a very recent theoretical study formulated on a square lattice by combining slave-particle analysis and quantum Monte Carlo \cite{PRB.99.245150},  the electronic entropy induced Mott gap reduction can also be one order of magnitude larger than $k_BT$, comparable to the lattice entropy effect quantified above. Therefore, the real finite-temperature Mott gap observed in experiment is expected to have an even steeper slope with respect to temperature. 

Heuristically, we expect that the experimentally observed Mott gap $\Delta_{Mott}$ gradually deviates downward from the DFT+U band gap as temperature increases.  A schematic illustration of the possible consequences is presented in Fig. \ref{transition}.  In principle, the gap melting can be momentum dependent \cite{cluster}. In an intermediate temperature range, the pseudogap state may emerge. Accrodingly, $\Delta_{Mott}$ is plotted with a finite width instead of a single-value curve.

Figure \ref{transition}(a) reflects a scenario without considering the lattice entropy effect. Given that the  CCDW-triggered MIT is first-order,  the order parmeters jump to some fixed values below the transition temperature. The dominating low-temperature dynamics comes from spin, which in addition renormalizes the Mott gap. 

Figure \ref{transition}(b) shows a trivial possibility, where lattice entropy simply further reduces the gap size. Exotic outcomes occur when the cooperation of electronic and lattice entropy melts the Mott gap before reaching the CCDW transtion [Figs. \ref{transition}(c,d)]. The lower and upper bounds of $\Delta_{Mott}$ define two additional charateristic temperatures, which we term as $T_{Mott}$ and $T^*$. Depending on the positions of $T_{Mott}$ and $T^*$ with respect to $T_{CCDW}$, the phase diagram can have a richer structure.  

A revisit of the experimental data suggests that it deserves further investiations to address the questions: Is it proper to assign the whole regime below $T_{CCDW}$ as a Mott insulator? Do additional characteristic temperatures exist? The scanning tunneling spectrocopy shows that while the energy splitting between the lower Hubbard peak and the upper Hubbard peak at 5 K \cite{qiao2017} and 78 K \cite{PRB_2015_cho} appear to fall on our $\langle E_g^{BO}(T) \rangle$ curve, some in-gap density of states has emerged at the elevated temperature. At 130 K \cite{PRB_2018_lutsyk}, the gap profile has transformed into the V shape.  On the other hand, the recent nuclear quadruple resonance measurement \cite{2017high} shows that below $T_{CCDW}$, the spin-lattice relaxation rate $1/T_1$ undergoes an anomalous transition from $T^2$ to a much steeper $T^4$ power law.  The transition temperature is decided as 55 K. Around this temperature, the in-plane resistivity was long noticed to undergo a crossover from a metallic behavior to an insulator behavior \cite{PMB_1979}. 

\section{Conclusion}

In summary, we predict a giant reduction of the Mott gap in 1T-TaS$_2$ induced by lattice vibrations.  The electronic entropy and spin fluctuations are expected to give rise to an even stronger temperature dependence, presenting 1T-TaS$_2$ a feasible experimental platform to observe a continuous thermal evolution of the Mott phase.  
It is also worth applying this computational methodology to other transition-metal dichalcogenides as well as oxides, to understand the general trend of lattice entropy effect in MIT.

\section{Acknowledgement}

We acknowledge Yayu Wang and Xintong Li for helpful discussion. This work is supported by NSFC under Grant Nos. 11774196, 11504040 and 11904350, Tsinghua University Initiative Scientific Research Program, Open Research Fund Program of the State Key Laboratory of Low-Dimensional Quantum Physics (No. KF201804), and the Fundamental Research Funds for the Central Universities of China (No. DUT16RC(4)66).

\bibliography{ref} 
\end{document}